\documentclass[conference]{IEEEtran}
\IEEEoverridecommandlockouts
\usepackage{cite}
\usepackage{amsmath,amssymb,amsfonts}
\usepackage{algorithmic}
\usepackage{graphicx}
\usepackage{subcaption}
\usepackage{textcomp}
\usepackage{enumerate}
\usepackage{xcolor}
\usepackage{hyperref}
\def\BibTeX{{\rm B\kern-.05em{\sc i\kern-.025em b}\kern-.08em
    T\kern-.1667em\lower.7ex\hbox{E}\kern-.125emX}}
\begin{document}

\title{Depthwise Separable CNN for D-MIMO Indoor Localization with Data Reduction}

\author{\IEEEauthorblockN{Georgios Mystriotis}
\IEEEauthorblockA{\textit{ESAT, WaveCoRE} \\
\textit{KU Leuven}\\
Leuven, Belgium \\
george.mystriotis@kuleuven.be}
\and
\IEEEauthorblockN{Dr. Rodney Martinez Alonso}
\IEEEauthorblockA{\textit{ESAT, WaveCoRE} \\
\textit{KU Leuven}\\
Leuven, Belgium \\
rodney.martinezalonso@kuleuven.be}
\and
\IEEEauthorblockN{Dr. Achiel Colpaert}
\IEEEauthorblockA{\textit{imec} \\
Kapeldreef 75\\ 3001 Leuven, Belgium \\
achiel.colpaert@imec.be}
\and
\IEEEauthorblockN{Prof. Sofie Pollin}
\IEEEauthorblockA{\textit{imec} \\
Kapeldreef 75\\ 3001 Leuven, Belgium \\
sofie.pollin@imec.be}
}

\maketitle

\begin{abstract}
Indoor localization using Distributed Multiple-Input Multiple-Output (D-MIMO) and machine learning (ML) achieves sub-centimeter accuracy but faces midhaul capacity bottlenecks when transmitting raw Channel State Information (CSI) in Open Radio Access Networks (O-RAN) architectures. To address this, we propose a lightweight, distributed ML framework that shifts initial processing to the network edge. By deploying localized models as dApps on Distributed Units (DUs), each requiring just 1.39 MB of memory and 1.96 MFLOPs, the system performs CSI feature extraction and reduction on the edge. The reduced low-dimensional features are transmitted to the Central Unit (CU), where another dApp is deployed for location estimation. Evaluated on a high-density dataset, this framework reduces midhaul traffic by 100x while maintaining an average error of 8.5 mm, even with half the deployed Radio Units (RUs), providing a scalable blueprint for practical D-MIMO localization.
\end{abstract}


\section{Introduction}
Driven by a growing demand for highly accurate position estimation, indoor localization has emerged as a critical research domain to bridge the navigational gap in GPS-denied environments \cite{9159543}. In complex indoor environments characterized by severe multipath fading and non-line-of-sight (NLoS) conditions, traditional positioning technologies often fall short. Recently, Distributed Multiple-Input Multiple-Output (D-MIMO) systems have emerged as an effective solution. When combined with Machine Learning (ML) fingerprinting techniques, which map complex spatial signatures to physical coordinates, D-MIMO networks can leverage massive spatial diversity to achieve centimeter-level localization accuracy \cite{zholamanov2025rssi}.  

Despite these advantages, the practical deployment of ML-assisted D-MIMO localization is hindered by architectural constraints. In traditional Open Radio Access Network (O-RAN) deployments, Radio Units (RUs) act merely as data collectors, transmitting high-dimensional raw frequency information to a Distributed Unit (DU) for processing \cite{11275403}. As the number of deployed antennas increases to improve spatial resolution, the volume of raw data grows drastically, requiring it to be transmitted to the Central Unit (CU). This centralized paradigm creates midhaul capacity bottlenecks, rendering the system unscalable for real-time applications or dense deployments \cite{zhu2024resource}.  

Although recent literature has begun exploring decentralized processing to mitigate these midhaul constraints \cite{zhu2024resource}, existing solutions often overlook the real-time processing constraints inherent to DUs. Moving standard Deep Neural Networks (DNNs) to the network edge shifts the burden to the DU hardware, leading to processing delays due to larger model size and a high number of FLOPs \cite{ngo2025edge}. Furthermore, to the best of our knowledge, there is a lack of research addressing the trade-offs among the number of RUs deployed, ML computational complexity, and overall localization accuracy. 

To overcome these bottlenecks, this paper proposes a novel, lightweight, distributed ML framework that shifts the initial CSI processing directly to the network edge. By deploying MobileNet-inspired models on the DUs, our system simultaneously performs feature extraction and reduction. Rather than forwarding raw CSI, the DUs transmit only low-dimensional features to the CU for the location estimation. 

The main contributions of this paper are summarized as follows:
\begin{enumerate}
    \item Lightweight edge processing: We introduce a lightweight, distributed ML architecture deployed directly at the DUs (1.39 MB, 1.96 MFLOPs per inference), enabling edge-based CSI feature extraction.
    \item Data Reduction: We demonstrate that our distributed model effectively reduces midhaul traffic by a factor of up to 100 compared to raw CSI transmission, while maintaining sub-centimeter accuracy. 
    \item RU-complexity tradeoff analysis: We evaluate the tradeoffs between the number of RUs deployed, model complexity, and accuracy, demonstrating that our framework achieves sub-centimeter accuracy even in scenarios with fewer RUs. 
\end{enumerate}

\section{Related Work}

\subsection{D-MIMO and ML-Assisted Indoor Localization}

Recent advances in indoor localization have significantly improved accuracy by integrating vision-based architectures into traditional pipelines. For example, Swin Transformer-based models have been successfully employed to mitigate CSI fingerprint distortion, significantly outperforming traditional DNNs, achieving accuracies around 8mm while also reducing fingerprint storage overhead \cite{xu2024swinloc}.

However, the majority of these highly accurate ML frameworks, including the latest Transformer architectures, still fundamentally rely on a centralized processing paradigm, assuming that all raw CSI can be seamlessly aggregated and processed in a single location.

\subsection{Lightweight Localization Fingerprinting at the Edge}

CSI fingerprinting remains the gold standard for centimeter-level indoor positioning due to its granular capture of amplitude and phase shifts \cite{s23135830}. Nevertheless, standard DNNs impose a significant computational
burden on network infrastructure, motivating a shift toward lightweight,
edge-native ML deployment \cite{martin2025fingerprinting}.

Recent advancements have addressed this need by designing neural network architectures tailored to mobile constraints. For example, CSILoc employs a two-channel 1D CNN on preprocessed CSI magnitude and phase data \cite{9918112}. By avoiding complex 2D convolutions and traditional pooling layers, this 1D CNN architecture requires a minimal memory footprint (11,280 to 12,600 parameters), while achieving 1.89 ms inference on commercial smartphones and improved localization performance \cite{9918112}.

Alternative approaches reduce feature dimensionality via vector embeddings adapted from speech processing, extracting d-vectors from a truncated DNN and classifying with a simpler backend such as time-reversal resonating strength (TRRS) \cite{REYES2025125802}. This approach requires only 82.01\% of the original DNN's parameter count while maintaining comparable localization accuracy, substantially reducing the storage space and processing power required by IoT nodes \cite{REYES2025125802}.

\subsection{ML-Assisted Data Reduction in O-RAN}

In standard O-RAN architectures, the limited capacity of fronthaul and midhaul links often creates transmission bottlenecks. Traditional compression techniques specified by the O-RAN architecture, such as Block Floating Point (BFP) and $\mu$-law companding, often struggle to balance the compression ratio with the signal fidelity required for more complex tasks \cite{11275403}.

Recent work has shifted toward AI-driven reduction, using distributed and
meta-learning to jointly optimize edge transformation and central decoding.
Neural transform-compress-forward schemes reduce uplink signal dimensionality
before transmission, outperforming mathematical compression and achieving up to
64:1 ratios \cite{10624764, zhu2024resource, 11271118}. However, these methods target generic throughput rather than preserving task-specific features (e.g., for localization), and degrade at high reduction ratios.

The LITE framework addresses this via compression-aware training with an 
autoencoder at the O-DU \cite{goez2026lite}. By compressing CSI signaling by 50\%, an attention-enhanced predictor at the Near-RT-RIC achieves a 4.6x throughput gain while drastically reducing model complexity. However, this 2x reduction factor remains insufficient for dense or heavily constrained networks.

\section{System Model}

\subsection{Network Architecture and Geometric Topology}
We consider a D-MIMO uplink localization scenario within an O-RAN architecture. The physical topology consists of $K$ distributed RUs deployed around the User Equipment (UE). These RUs enclose a predefined square planar Region of Interest (RoI), denoted as $\mathcal{A}\in\mathbb{R}^3$, in which a single-antenna UE operates. 

\begin{figure}[htbp]
\centerline{\includegraphics[scale=0.26]{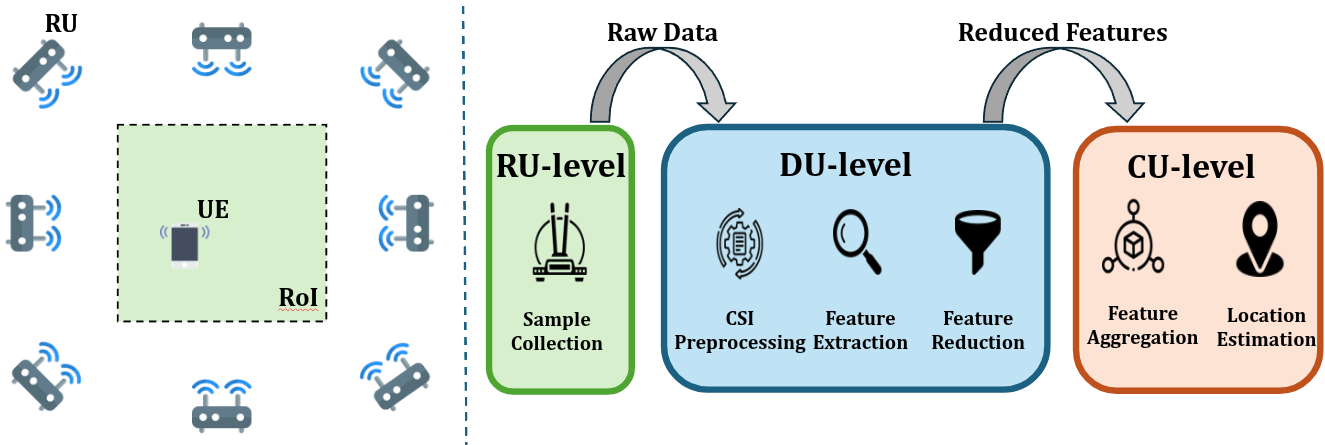}}
\caption{System Topology \& Architecture.}
\label{fig1}
\end{figure}

Each RU is equipped with an $N$-element Uniform Linear Array (ULA), where $N = 8$. The $K$ distributed RUs are connected to $M$ DUs that subsequently connect to a single CU (Fig. \ref{fig1}).

\subsection{Signal Model and CSI Acquisition (RU-Level)}
Let the true spatial coordinates of the UE be denoted as $\mathbf{p} = [x, y, z]^T$, inside the square operating area ($\mathbf{p} \in \mathcal{A}$). To perform the localization, the UE transmits standard uplink pilot signals. The estimated CSI matrix at the $k$-th DU, denoted as $\mathbf{H}_k \in \mathbb{C}^{N \times S}$, with $S$ the number of subcarriers, implicitly captures the spatial signatures of the multipath channel.

\subsection{Edge Feature Extraction and Data Reduction (DU-Level)}
To alleviate midhaul link capacity constraints, feature extraction and data reduction are performed directly at the edge. Instead of forwarding raw CSI matrices, the $k$-th DU utilizes a local feature extraction neural network to map the high-dimensional CSI to a lower-dimensional latent representation:
\begin{equation}
    \mathbf{Z}_k = f_{\theta_{RU}}(\mathbf{H}_k, \eta),
\end{equation}
where $f_{\theta_{DU}}(\cdot)$ represents the feature extraction function parameterized by $\theta_{DU}$. This process reduces the data by a predefined reduction ratio $\eta$, retaining only the most significant features necessary for localization. The reduced feature tensors $\mathbf{Z}_k$ are then transmitted to the CU over the midhaul interface.

\subsection{Feature Aggregation and Centralized Fusion (CU-Level)}
The CU acts as both an aggregator and a fuser of the information to predict the UE's coordinates. The CU collects the reduced features from all the DU and aggregates them to reconstruct the global spatial context:

\begin{equation}
    \mathbf{Z}_{global} = g_{aggregate}(\mathbf{Z}_1, \mathbf{Z}_2, \dots, \mathbf{Z}_K).
\end{equation}
Finally, this global feature map is passed through a centralized localization neural network running at the CU to output the estimated coordinates of the UE, $\mathbf{\hat{p}}$:

\begin{equation}
    \mathbf{\hat{p}} = h_{\theta_{CU}}(\mathbf{Z}_{global}),
\end{equation}
where $h_{\theta_{CU}}(\cdot)$ represents the regression network parameterized by $\theta_{CU}$.

\section{Proposed ML Framework}

\subsection{Data Preprocessing}

To ensure the neural network can effectively capture the underlying physical characteristics of the wireless channel, the raw complex-valued CSI is preprocessed to enrich the channel representation.

\subsubsection{Data Reshaping and Real-Imaginary Separation} 
The raw CSI data initially exists as a complex-valued array. Because standard deep learning frameworks are optimized for real-valued arithmetic, the complex CSI matrices are first transformed. A complex matrix of size $S \times A$ (where $S$ represents the number of subcarriers and $A$ represents the number of antennas) is split into its real and imaginary components. The data is then reshaped and stacked along a new dimension, resulting in samples of shape $2 \times S \times A$.

\subsubsection{Multi-Domain Feature Extraction} 
Based on \cite{9129126} we extract complementary representations of the channel in our dApp preprocessing pipeline (Fig. \ref{fig1}). Rather than feeding only the Cartesian representation into the neural network, this multi-domain approach explicitly exposes amplitude, phase, delay, and angular characteristics to the neural network. For a given input tensor, three distinct feature transformations are applied:

\begin{itemize}
    \item \textit{Polar Conversion:} The model computes the amplitude and phase for each CSI element.

    \item \textit{Time-Domain Conversion (Delay Profile):} To capture the multipath delay characteristics of the channel, an Inverse Fast Fourier Transform (IFFT) is applied along the subcarrier dimension ($S$).

    \item \textit{Angle-Delay Domain Conversion (2D FFT):} To capture the spatial-frequency characteristics, a 2-dimensional Fast Fourier Transform (2D FFT) is applied across both the subcarrier ($S$) and antenna ($A$) dimensions. 
\end{itemize}

\subsubsection{Feature Concatenation} 
In the final step of preprocessing, the tensors generated by the three transformations are concatenated, yielding samples of shape $6 \times S \times A$. By explicitly providing these physical features, the burden on the initial layers of the network to discover these transformations is reduced, facilitating faster convergence and improved representation learning.

\subsection{DU Feature Extraction and Data Reduction}

In the proposed D-MIMO O-RAN architecture, each DU must process and transmit channel observations to the CU over capacity-constrained midhaul links. To achieve sub-centimeter localization accuracy under midhaul constraints, we employ DU-hosted dApps that leverage lightweight neural networks for CSI feature extraction and data reduction (Fig. \ref{fig1}). The model is specifically designed for edge deployment, utilizing a MobileNet-inspired architecture with Depthwise-Separable Convolutions (DSCs) and channel-attention mechanisms to minimize computational complexity \cite{Sandler_2018_CVPR}. The key differences lie in the use of 1D instead of 2D convolutions across the subcarrier dimension to further reduce complexity, the adoption of the smoother GELU activation over ReLU, and the integration of a Channel-Attentive Bottleneck Head.

\begin{figure}[htbp]
\centerline{\includegraphics[scale=0.25]{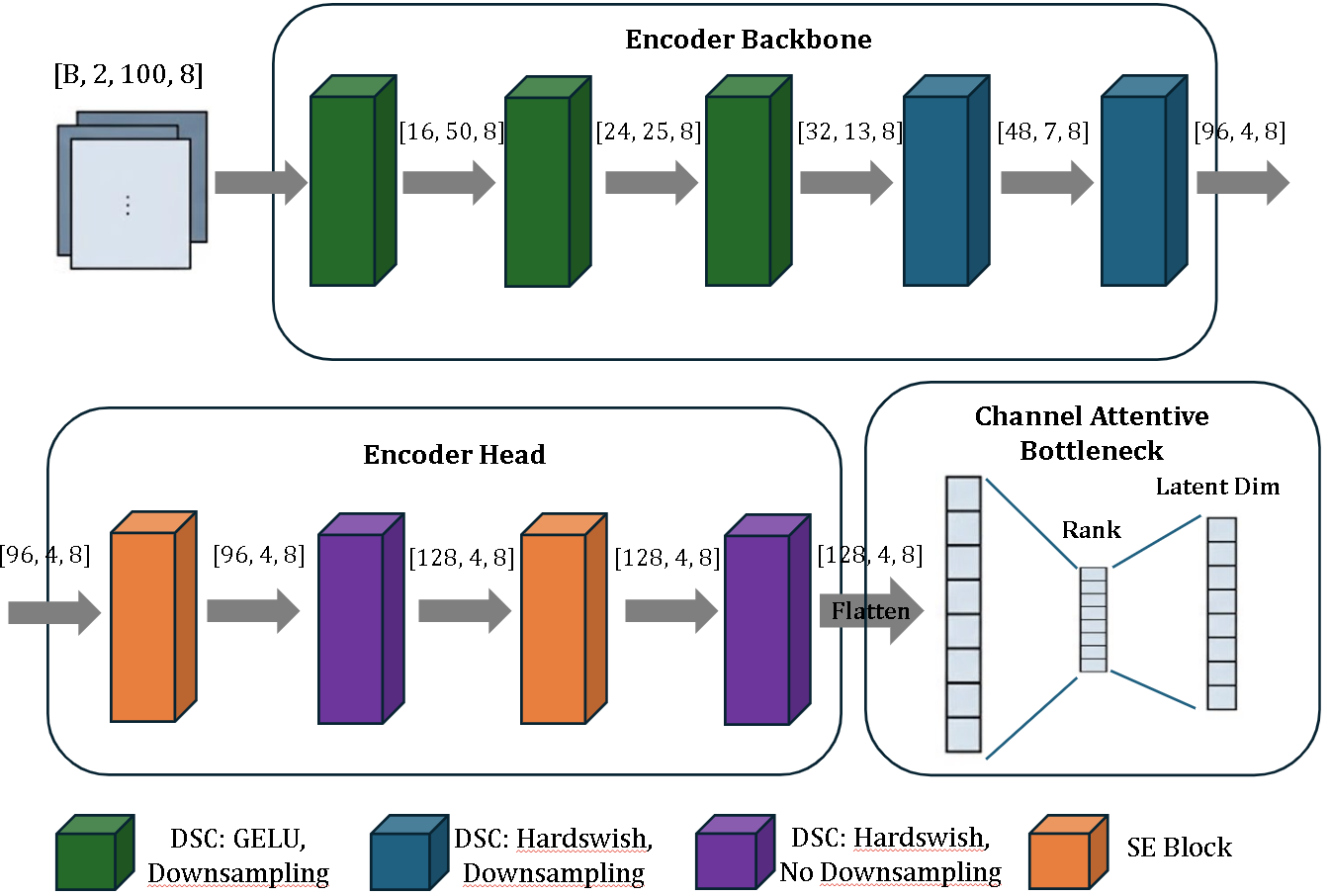}}
\caption{Distributed Model Architecture}
\label{model_architecture}
\end{figure}

The DU model consists of three primary stages: a feature-extraction backbone, an intermediate adapter, and a channel-attentive bottleneck-reduction head. 

\subsubsection{Lightweight Feature Extraction Backbone}
Let the input channel tensor at the DU be denoted as $\mathbf{X} \in \mathbb{R}^{F \times S \times A}$, where $F$ is the number of input channels, and $S$ and $A$ represent the number of subcarriers and number of antennas, respectively. To efficiently extract localized features, the backbone processes $\mathbf{X}$ through five DSC blocks. 

Unlike standard convolutions, each DSC block factorizes the operation into a depthwise spatial convolution and a $1 \times 1$ pointwise convolution. This significantly reduces the parameter count and computational footprint \cite{Sandler_2018_CVPR}. To manage the scale of the features, we employ Group Normalization paired with non-linear activation functions, specifically GELU in the early layers and Hardswish in the deeper layers \cite{Sandler_2018_CVPR}. Spatial downsampling (across the subcarrier dimension) is applied to each block, serving as a learnable pooling layer that increases the model's expressiveness while reducing computational load in deeper layers.

\subsubsection{Adapter Head and Squeeze-and-Excitation Recalibration}
To refine the semantic features before reduction, the backbone output is passed through an adapter module. This module interleaves additional DSC blocks with Squeeze-and-Excitation (SE) blocks. The SE blocks explicitly model channel interdependencies by dynamically recalibrating channel-wise feature responses via a Hardsigmoid gating mechanism. This ensures that the network prioritizes the most informative features for the downstream localization task. 

\subsubsection{Channel-Attentive Bottleneck Head}
The final stage is responsible for aggregating the refined features and reducing them into a latent vector representation $\mathbf{z} \in \mathbb{R}^{l}$, where $l$ is the dimension of the latent vector, strictly determined by the predefined data reduction ratio $\eta$. 

Let the output of the adapter be $\mathbf{X'} \in \mathbb{R}^{F' \times S' \times A'}$. The bottleneck head first computes a channel descriptor $\mathbf{g} \in \mathbb{R}^{F'}$ using global average pooling:
\begin{equation}
    \mathbf{g}_c = \frac{1}{S' A'} \sum_{i=1}^{S'} \sum_{j=1}^{A'} \mathbf{X'}_{f,i,j}.
\end{equation}

This descriptor is fed into a small multi-layer perceptron (MLP) with a Hardsigmoid activation to generate channel attention logits, $\mathbf{a} \in [0, 1]^{F'}$. These logits are broadcast and applied element-wise to the feature map to produce the attention-weighted features $\tilde{\mathbf{X'}}$:
\begin{equation}
    \tilde{\mathbf{X'}} = \mathbf{X'} \odot \mathbf{a},
\end{equation}
where $\odot$ denotes channel-wise multiplication.

Finally, $\tilde{\mathbf{X'}}$ is flattened and passed through a projection bottleneck to achieve the desired reduction. The bottleneck utilizes a low-rank intermediate projection to further constrain complexity. The resulting projected latent tensor $\mathbf{z}$ is then transmitted over the midhaul interface.

\subsection{Centralized Aggregation and Localization at the CU}

Upon receiving the reduced data from the DUs, the CU aggregates them and performs coordinate regression. Let $\mathbf{z}_k \in \mathbb{R}^{d_{latent}}$ represent the latent feature tensor received from the $k$-th DU.

Since the distributed RUs observe the UE from different spatial vantage points, the features extracted from their corresponding DUs must be fused to exploit the diversity of the D-MIMO system. In this architecture, we employ a direct concatenation strategy, aggregating everything into a single global feature vector $\mathbf{z}_{global} \in \mathbb{R}^{K \cdot d_{latent}}$:
\begin{equation}
    \mathbf{z}_{global} = \left[ \mathbf{z}_1^\top, \mathbf{z}_2^\top, \dots, \mathbf{z}_K^\top \right]^\top
\end{equation}

The aggregated global feature vector $\mathbf{z}_{global}$ serves as input to a centralized Feed-Forward Neural Network (FFNN) that predicts the UE's 3D coordinates. 

\section{Results}

To assess the performance of the proposed distributed framework, we utilize the ``Ultra-dense indoor MaMIMO CSI dataset'' \cite{nr6k-8r78-21}. Specifically, the ML model is evaluated using the DIS LoS scenario, which provides dense CSI fingerprints in an environment for D-MIMO systems. 

\subsection{Localization Accuracy and RU deployment Tradeoffs}
\begin{figure}[htbp]
    \centering
    \includegraphics[width=\linewidth]{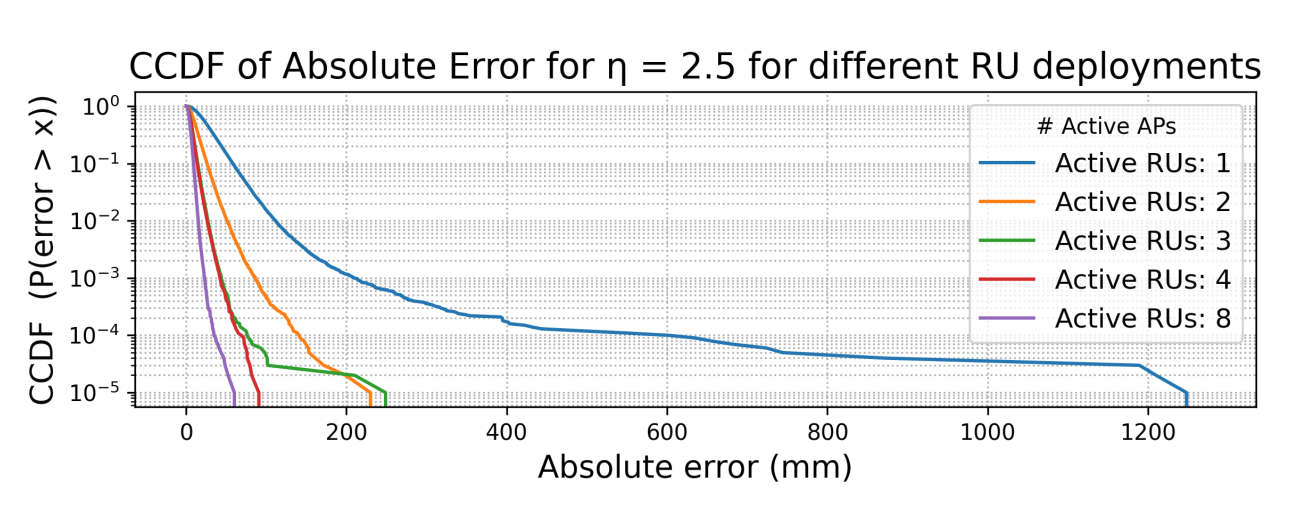}
    \caption{CCDF of absolute positioning error for $\eta = 2.5$ across different active RU configurations.}
    \label{fig:ccdf_accuracy}
\end{figure}

One objective of this framework is to understand the tradeoff between the number of deployed RUs and the resulting localization accuracy. Fig. \ref{fig:ccdf_accuracy} illustrates the Complementary Cumulative Distribution Function (CCDF) of the absolute positioning error at a baseline reduction ratio of 2.5. 

The results demonstrate that while deploying all 8 RUs yields the highest accuracy ($\approx 5.3 mm$), it is not strictly necessary to achieve sub-centimeter-level accuracy. Remarkably, reducing the active RUs to 4 or even 3 still yields subcentimeter average localization errors ($\approx 7.5 mm$ and $\approx 8 mm$, respectively). This confirms that our end-to-end model captures the spatial signatures required for accurate positioning while relaxing RU deployment requirements.

\subsection{Computational Complexity and Edge Feasibility}
\begin{figure}[htbp]
    \centering
    \hspace*{-0.5cm}
    \begin{subfigure}{0.51\linewidth}
        \centering
        \includegraphics[width=0.9\linewidth]{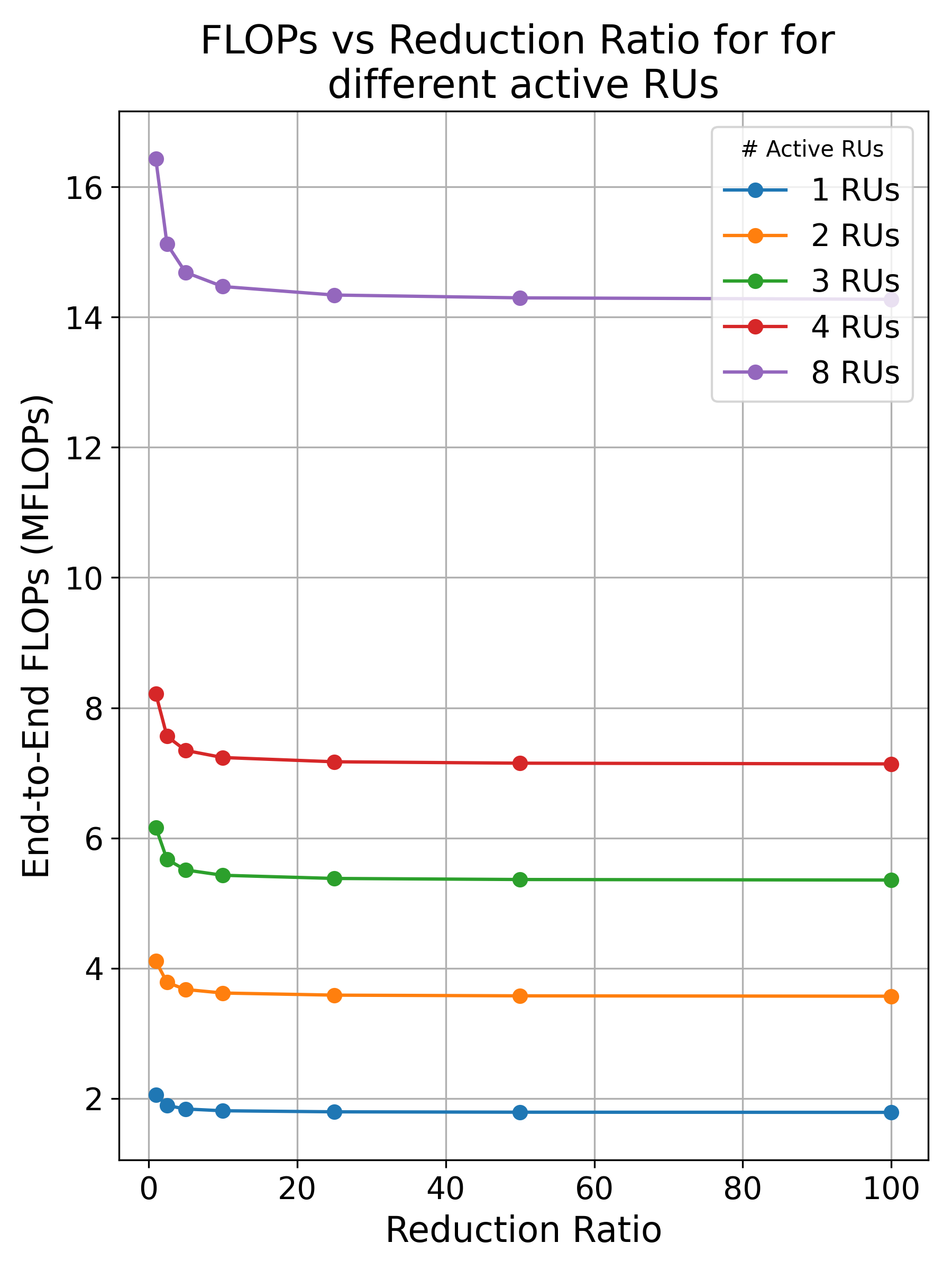} 
        \caption{End-to-End FLOPs.}
        \label{fig:complexity_flops}
    \end{subfigure}
    \begin{subfigure}{0.51\linewidth}
        \centering
        \includegraphics[width=0.9\linewidth]{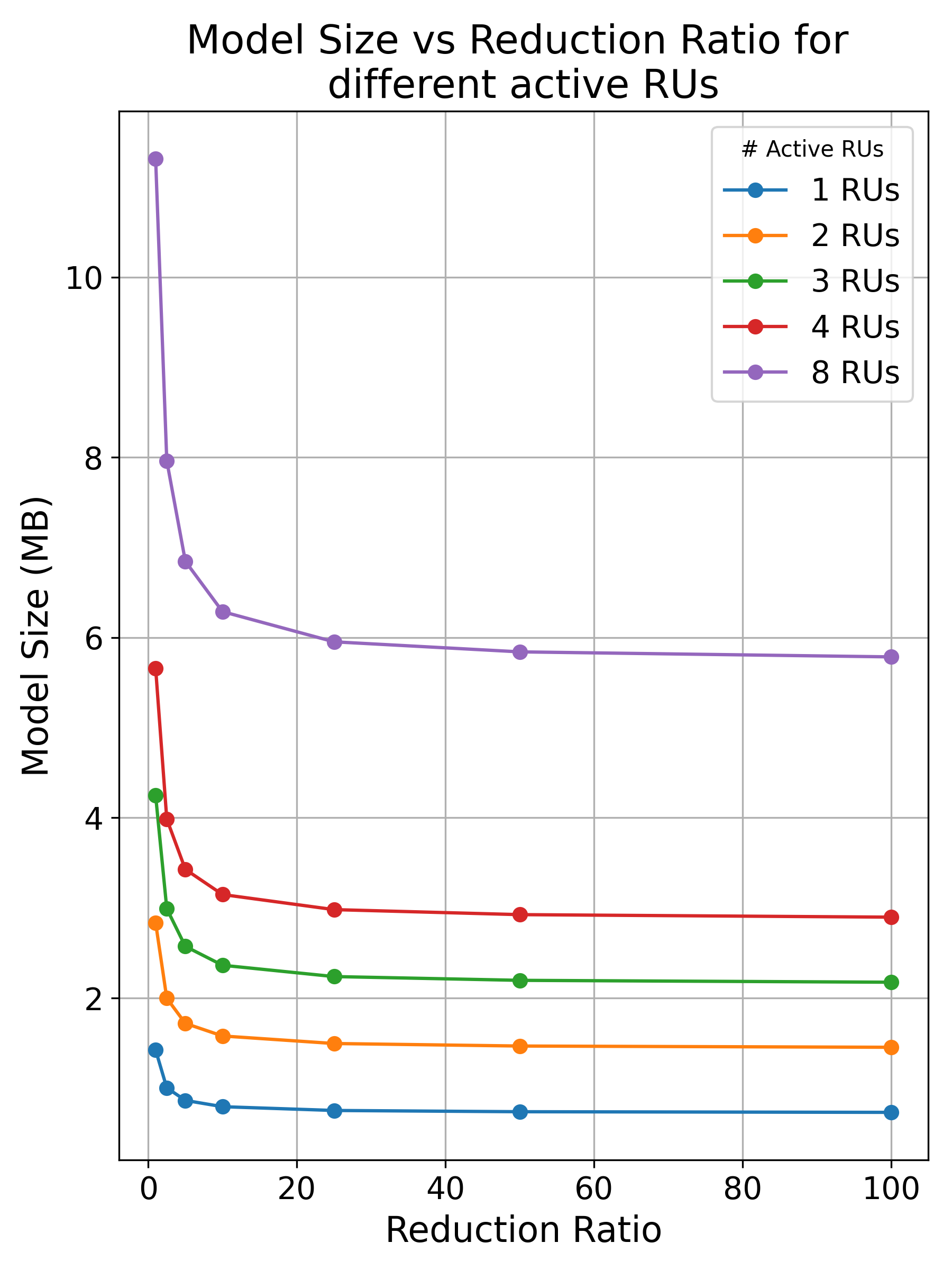} 
        \caption{Model Size (MB).}
        \label{fig:complexity_size}
    \end{subfigure}
    
    \caption{Computational complexity and memory footprint across reduction ratios for varying numbers of active RUs.}
    \label{fig:complexity_combined}
\end{figure}

To verify that our dApps can be practically deployed in a live O-RAN network, we evaluated the computational complexity and memory footprint of the framework across different RU deployments and reduction ratios (Fig. \ref{fig:complexity_combined}a and Fig. \ref{fig:complexity_combined}b). 

The results indicate that the end-to-end model complexity is primarily reduced by decreasing the number of active RUs (and consequently, the number of distributed models), rather than relying solely on the reduction ratio. Each distributed RU model operates with a minimal memory footprint of approximately 1.39 MB and executes just 1.96 MFLOPs at a reduction ratio of 1, even without quantization. By moving away from the high number of 2D convolution channels (e.g., 256 per layer) typical of previous approaches \cite{zhu2024resource}, we achieve around a 500-fold reduction in computational complexity. The reduction ratio mostly affects the computational complexity of the Channel-Attentive Bottleneck, which is already optimized by utilizing a low-rank intermediate projection, thus we do not see much improvement in complexity as we increase the reduction ratio.

\subsection{Data Reduction Capabilities}
To assess the data reduction capabilities, we compared three distinct deployment scenarios: 1) The full 64-antenna setup with $\eta = 2.5$, 2) a 32-antenna setup with $\eta = 2.5$, and 3) a 32-antenna setup with $\eta = 100$.

\begin{figure}[htbp]
    \centering
    \includegraphics[width=\linewidth]{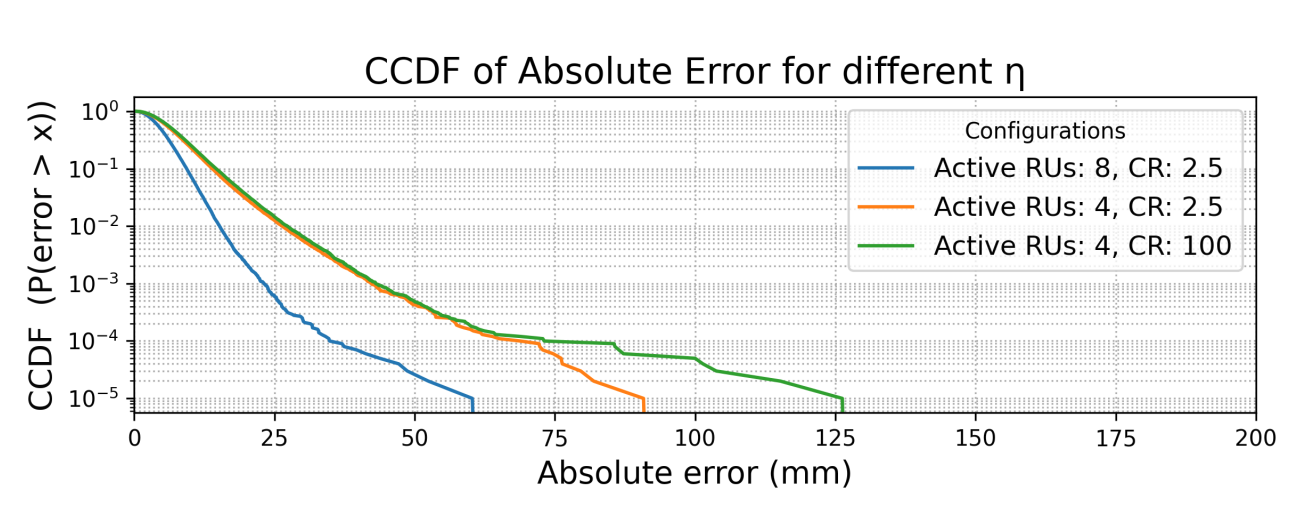}
    \caption{CCDF of absolute positioning error across different Reduction Ratios \& active RU configurations.}
    \label{fig:ccdf_cr_comparison}
\end{figure}

As shown in \ref{fig:ccdf_cr_comparison}, even when increasing the reduction ratio to 100 in the 32-antenna topology, the framework successfully maintains centimeter-level accuracy ($\approx 8.5 mm$). This demonstrates that our distributed models not only act as efficient feature extractors but are also capable of accurate
dimensionality reduction.

\section{Conclusion}
In this paper, we introduced a novel, lightweight, distributed Machine Learning framework designed to address the midhaul capacity and computational load bottlenecks inherent in centralized D-MIMO indoor localization systems. By shifting the initial CSI processing directly to the network edge, our architecture enables DUs to simultaneously execute CSI feature extraction and data reduction.
Our evaluations, conducted on a high-density fingerprinting dataset, demonstrated several advantages: 1) the DU models are exceptionally lightweight, requiring a minimal memory footprint of just 1.39 MB and executing only 1.96 MFLOPs per model, remaining well within the real-time execution windows required by the DU, 2) the proposed framework effectively reduces midhaul data traffic by a factor of up to 100, without compromising accuracy and 3) the system manages to maintain centimeter-level accuracy even when half of the RUs are deployed.
By drastically reducing bandwidth requirements without sacrificing accuracy, this framework provides a scalable, computationally efficient blueprint for the practical deployment of D-MIMO indoor localization networks within O-RAN distributed architectures.
The code to reproduce our simulations and experiments is available on our \href{https://gitlab.kuleuven.be/networked-systems/public/dscnn-localization}{\underline{gitlab}}.

\section*{Acknowledgement}
This paper was supported by the SUNRISE-6G project, which has received funding from the European Union under Grant Agreement No. 101139257. Views and opinions expressed are, however, those of the authors only and do not necessarily reflect those of the European Union or the Smart Networks and Services Joint Undertaking. The work of Rodney Martinez Alonso is supported by the Research Foundation–Flanders (FWO) under Grant 1211926N. This research received funding from the Flemish Government under the “Onderzoeksprogramma Artificiële Intelligentie (AI) Vlaanderen” programme.

\bibliographystyle{ieeetr} 
\bibliography{references}  

\end{document}